\documentclass[preprint2]{aastex}
\usepackage{amsmath}
\usepackage[usenames,dvipsnames]{xcolor}

\shorttitle{The $M_{\rm BH}-\sigma_{\rm rms}^2$ relation in the low-mass Regime}
\shortauthors{Pan et al.}

\begin{document}


\title{On the Black Hole Mass---X-ray Excess Variance Scaling Relation for Active Galactic Nuclei in the Low-mass Regime}


\author{Hai-Wu Pan\altaffilmark{1,2}, Weimin Yuan\altaffilmark{1,2}, Xin-Lin Zhou\altaffilmark{1,2}, Xiao-Bo Dong\altaffilmark{3} \& Bifang Liu\altaffilmark{1,2}}

\altaffiltext{1}{Key Laboratory for Space Astronomy and Technology, Chinese Academy of Sciences, China; panhaiwu@nao.cas.cn, wmy@nao.cas.cn}
\altaffiltext{2}{National Astronomical Observatories, Chinese Academy of Sciences, 20A Datun Road, Chaoyang District, Beijing, China}
\altaffiltext{3}{Yunnan Observatories, Chinese Academy of Sciences, Kunming, Yunnan, China}



\begin{abstract}
Recent studies of active galactic nuclei (AGN) found a statistical inverse linear scaling between the X-ray normalized excess variance $\sigma_{\rm rms}^2$ (variability amplitude) and the black hole mass spanning over $M_{\rm BH}=10^6- 10^9\ M_{\odot}$. Being suggested to have a small scatter, this scaling relation may provide a novel method to estimate the black hole mass of AGN. However, a question arises as to whether this relation can be extended to the low-mass regime below $\sim10^6\ M_{\odot}$. If confirmed, it would provide an efficient tool to search for AGN with low-mass black holes using X-ray variability. This paper presents a study of the X-ray excess variances for a sample of AGN with black hole masses in the range of $10^5- 10^6\ M_{\odot}$ observed with {\it XMM-Newton} and {\it ROSAT}, including data both from the archives and from newly preformed observations. It is found that the relation is no longer a simple extrapolation of the linear scaling; instead, the relation starts to flatten at $\sim10^6\ M_{\odot}$ toward lower masses. Our result is consistent with the recent finding of \citet{L15}. Such a flattening of the $M_{\rm BH}-\sigma_{\rm rms}^2$ relation is actually expected from the shape of the power spectrum density of AGN, whose break frequency is inversely scaled with the mass of black holes.
\end{abstract}


\section{INTRODUCTION}

Rapid X-ray variability is one of the basic observational characteristics of active galactic nuclei (AGN) \citep{M85,G92,M93}. It is a useful tool to study black holes (BH) and the central engine of AGN, since the X-ray emission is thought to originate from the innermost region of an accretion flow around the BH.
One commonly used method to characterize the variability is the power spectrum density (PSD) analysis, which quantifies the amount of variability power as a function of temporal frequency \citep{G93,L93}.
The PSD of AGN has been found to be well described by a broken power-law (e.g. \citealp{P95,E99,U02,M03,V03b}). The break frequency is found to be inversely scaled with the black hole mass in a linear way, with a possible dependence on the scaled accretion rate $\dot{m}$ (in units of the Eddington accretion rate) \citep{M06,G12}. These results are remarkable in the sense that AGN show similar X-ray variability properties to black hole X-ray binaries (BHXB), indicating that AGN are scaled-up versions of BHXB (see also \citealp{Z15B}).

However, reliable PSD analyses require well sampled, high quality X-ray data of time series, which are generally hard to obtain for large samples of AGN for the current X-ray observatories. Instead, an easier-to-calculate quantity, the ``normalized excess variance'' $\sigma_{\rm rms}^2$ (e.g. \citealp{N97,T99}), is commonly used to quantify the X-ray variability amplitude. Early studies revealed correlations between the excess variance and various parameters of AGN, such as X-ray luminosity, spectral index, and the FWHM of the H$\beta$ line (e.g. \citealp{N97,G00,T99,M01}). Later work suggested that these correlations are in fact by-products of a more fundamental relation with black hole mass: the $M_{\rm BH}-\sigma_{\rm rms}^2$ relation \citep{L01,B03,P04}. In fact, this relation conforms to the scaling relation for the PSD break frequency with BH mass \citep{P04,O05,P12}.
\citet{O05} confirmed the anti-correlation between the excess variance and $M_{\rm BH}$ with a large AGN sample observed with ASCA. \citet{Z10} obtained a tight correlation, using high quality {\it XMM-Newton} light curves of AGN whose black hole masses were measured with the reverberation mapping technique. The intrinsic dispersion of the relationship ($\sim 0.2$ dex) is comparable to that of the relation between $M_{\rm BH}$ and stellar velocity dispersion for galactic bulge \citep{T02}. By making use of a large sample of 161 AGN observed with {\it XMM-Newton} for at least 10\,ks for each object, \citet{P12} (henceforth P12) reaffirmed this relationship (using the excess variance calculated on various timescales of 10\,ks, 20\,ks, 40\,ks and 80\,ks), but found only a weak dependence on the accretion rate. The significant correlation with small scatters (0.4 dex for the reverberation mapping sample, 0.7 dex for the CAIXAvar sample, see \citealp{P12} for details) suggested that it may provide similarly or even more accurate black hole mass estimation compared to the method based on the single epoch optical spectra \citep{K00,V06}. Moreover, unlike the commonly used virial method that is susceptible to the orientation effect of AGN \citep{C04}, X-ray variability can be considered as inclination-independent.

However, the previous studies were based on AGN samples with mostly supermassive black hole of $M_{\rm BH}>10^6 \ M_{\odot}$, and little is known about the relation for AGN with $M_{\rm BH}<10^6 \ M_{\odot}$. The $\sigma_{\rm rms}^2$ of a few AGN with $M_{\rm BH} \sim 10^6 \ M_{\odot}$ were presented and found to have the largest $\sigma_{\rm rms}^2$ values among AGN over a large $M_{\rm BH}$ range \citep{M09,A11}. \citet{P12} proposed that the $M_{\rm BH}-\sigma_{\rm rms}^2$ relation may show a deviation from the linear relation in the low-mass regime; however, the data is too sparse to draw a firm conclusion. Thus, the question remains unanswered as to whether this relation can be extended to $M_{\rm BH}<10^6\ M_{\odot}$ \footnote{While we were writing this paper, a new paper \citep{L15} appeared very recently, which carried out a similar study and achieved a similar result as ours in this work.}.

The answer to this question is important in at least two aspects. First, if the answer is ``yes'', the relation would provide a valuable method to find the so-called low-mass AGN with $M_{\rm BH}\lesssim10^6\ M_{\odot}$, or sometimes referred to as AGN with intermediate-mass black holes (IMBH). This is of particular interest since in these AGN the commonly used virial method involving optical broad emission lines becomes difficult in practice due to the faint AGN (broad line) luminosities that are outshined by the host galaxy starlight. In fact, there has been a few attempts in practice by adopting such an assumption. By extrapolating the $M_{\rm BH}-\sigma_{\rm rms}^2$ relation to below $10^6\ M_{\odot}$, \citet{K12} selected a sample of 15 candidate low-mass AGN using the X-ray excess variance. Second, the $M_{\rm BH}-\sigma_{\rm rms}^2$ relation must have a cutoff somewhere, otherwise the variability would have become unrealistically large for ultra-luminous X-ray sources (ULX) and BHXB, which is not seen in observations, however (e.g. \citealp{G11,Z15}). From a theoretical perspective, the $M_{\rm BH}-\sigma_{\rm rms}^2$ relation is actually a manifestation of the inverse scaling of the break frequency of the AGN PSD with the black hole mass, since the excess variance is the integral of the PSD over frequency domain \citep{V89, V97}. In fact, a break of the $M_{\rm BH}-\sigma_{\rm rms}^2$ relation at low BH masses had been predicted based on the current understanding of the PSD of AGN \citep{P04,O05,P12}, and the exact break mass (the mass at which the break occurs) depends on the shape of the PSD. Therefore, a study of the $M_{\rm BH}-\sigma_{\rm rms}^2$ relation in the low-mass regime may provide a constraint on the break mass, as well as the shape of AGN PSD.

In this paper, we study the $M_{\rm BH}-\sigma_{\rm rms}^2$ relation in the $M_{\rm BH}=10^5- 10^6\ M_{\odot}$ range, using an optically selected low-mass AGN sample from our previous work \citep{D12}. We use both new observations and archival data obtained with {\it XMM-Newton}, as well as archival data of {\it ROSAT}.
This paper is organized as follows. The introduction of our sample and data reduction are presented in section 2. In section 3, the excess variance is introduced, as well as the PSD models of AGN concerned in this study. The results and discussion are presented in section 4. Throughout the paper, a cosmology with $H_{\rm 0}=70 \rm km s^{-1} Mpc^{-1}$, $\Omega_{\rm m}=0.3$ and $\Omega_{\rm \Lambda}=0.7$ is adopted.

\section{SAMPLE, OBSERVATION AND DATA REDUCTION}


There are over 300 low-mass (type 1) AGN with $M_{\rm BH} \lesssim 10^6\ M_{\odot}$ known so far. The largest is from our work \citep{D12}, which was selected homogeneously from the Sloan Digital Sky Survey (DR4), and comprises 309 objects with $M_{\rm BH} < 2\times10^6\ M_{\odot}$. The BH masses were estimated from the luminosity and the width of the broad H$\alpha$ line, using the virial mass formalism of \citet{G05,G07}. One feature of this sample is the accurate  measurements of the AGN spectral parameters, and hence the black hole masses and Eddington ratios. Compared to previous samples, this sample is more complete as it includes more objects with low Eddington ratios down to $L_{\rm bol}/L_{\rm Edd} \sim 0.01$ (see also \citealp{Y14}). We compile a working sample of low-mass AGN with usable X-ray data from this parent sample.

We search for X-ray observations from both the {\it XMM-Newton} and {\it ROSAT}  PSPC data archives to maximize the sample size. For {\it XMM-Newton} observations the 3XMM-DR4 catalogue is used. We also add new observations of three objects from our programme to study low-mass AGN with {\it XMM-Newton} (proposal ID: 074422, PI: W. Yuan). We consider only observations with exposure time longer than 10\,ks. There are 26 {\it XMM-Newton} observations for 16 objects and 6 {\it ROSAT} observations for 6 objects found in the archives. The new observations of J0914+0853, J1347+4743, and J1153+4612 were performed by {\it XMM-Newton} at faint imaging mode on November 1st, 22nd, and December 4th 2014 with an exposure time of 36\,ks, 31\,ks, and 14\,ks, respectively.

\begin{figure}
\epsscale{.80}
\plotone{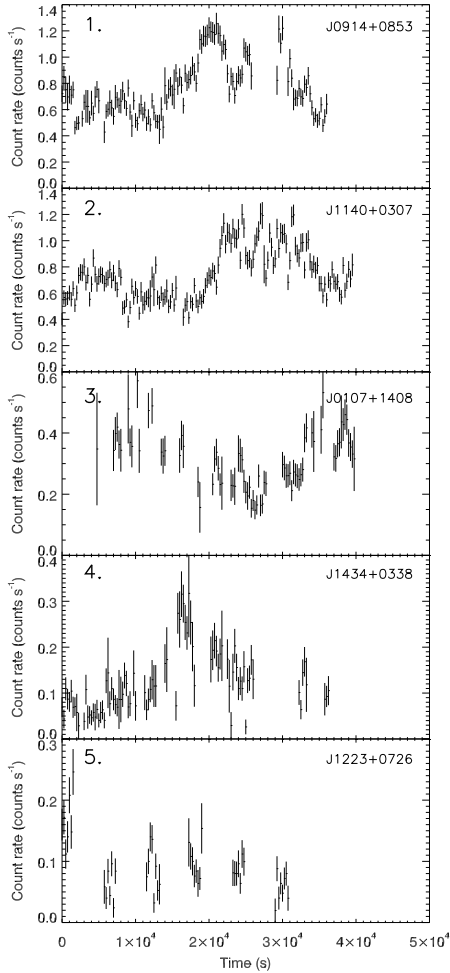}
\caption{Example light curves of low-mass AGN in our sample observed with {\it XMM-newton} (panels 1-4) and {\it ROSAT} (panel 5), respectively. All the light curves are extracted with 250\,s time bins.\label{fig1}}
\end{figure}

The X-ray data are retrieved from the {\it XMM-Newton} and {\it ROSAT} data archives. We follow the standard procedure for data reduction and analysis.
For the {\it XMM-Newton} observations, we use the data from the EPIC PN camera only, which have the highest signal-to-noise.
Light curves are extracted from observation data files (ODFs) by using the {\it XMM-Newton} Science Analysis System (SAS) version 12.0.1. Events in the periods of high flaring backgrounds are filtered out \footnote{Except the object J1153+4612, since it is bright enough (with a mean count rate=2.1 counts\,s$^{-1}$) that the aforementioned influence on timing analysis can be ignored.}. Observations with cleaned exposure time shorter than 10\,ks are also excluded. Typical source extraction regions are circles with a 40 arcsec radius. Only good events (single and double pixel events, i.e. PATTERN $\leq $ 4) are used for the PN data.
Background light curves are extracted from source-free circles with the same radius. Finally the SAS task EPICLCCORR is applied to make corrections for each of the {\it XMM-Newton} light curves. The energy band 0.2-10\,keV is used. The time bins of the light curves are chosen to be 250\,s, which are the same as in P12 for easy comparison. As demonstration, Figure \ref{fig1} shows some of the typical light curves (panels 1-4).

For {\it ROSAT} PSPC observations, the XSELECT package is used to extract source counts and light curves. Typical source extraction regions are circles of 50 arcsec radius, and the same aperture is used to extract background light curves. The energy band for {\it ROSAT} observations is 0.1-2.4\,keV. The time bins are also set to be 250\,s for the same reason. Examples of the {\it ROSAT} light curves are also shown in Figure \ref{fig1} (panel 5).

\begin{figure}[h]
\epsscale{1.20}
\plotone{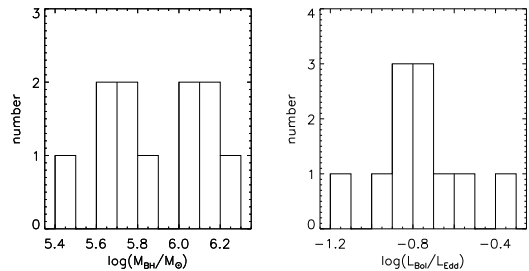}
\caption{Distributios of the black hole mass and scaled accretion rate for our sample, taken from \citet{D12}.\label{fig2}}
\end{figure}

All the light curves exhibit significant variability on short timescales, as shown in the figure.
For a few sources, due to short observational interval and relatively low signal-to-noise ratio (S/N), the intrinsic variability is overwhelmed by random fluctuations because of the large statistical uncertainties (the statistical uncertainty is larger than the source variability; see Section 3.1 for the definition of the excess variance). In such cases a meaningful excess variance cannot be obtained and its value is consistent with being zero (The same situation also happened in some objects or observations in previous studies, e.g. \citealp{O05,P12}). We find that such sources mostly have the mean S/N \footnote{The mean S/N is calculated as the ratio of the mean source count rate to its mean statistical error.} below 2.3. We thus introduce a cutoff on the mean S/N, below which the sources are dropped \footnote{One object with a signal to noise ratio (S/N$\sim$2.5) above the threshold (J1720+5748, observed with {\it ROSAT}) turns out to have the intrinsic variability consistent with zero. This is likely due to the stochastic nature of the X-ray variability of AGN, which happens to result in little variations within a relatively short time span. This object is also excluded.}.
Our final sample after the S/N cut is composed of 11 objects with 15 observations. Among them, 10 objects were observed with {\it XMM-Newton} for a total of 13 observations, and 2 objects observed with {\it ROSAT} for 2 observations (J1223+0726 was observed both in {\it XMM-Newton} and {\it ROSAT} observations).
Table \ref{tbl1} summarizes the basic parameters of the sample sources and the information on the X-ray observations. The black hole masses are taken from \citet{D12}, which are in the $10^5 - 10^6\ M_{\odot}$ range and the accretion rates in the $0.06 - 0.90$ range (Figure \ref{fig2}). All the objects are all at very low redshifts $z \leq 0.21$ with a median $z=0.090$.

\begin{deluxetable}{cccccccccc}
\tabletypesize{\scriptsize}
\rotate
\tablecaption{Sample and information on X-ray observations \label{tbl1}}
\tablewidth{0pt}
\tablehead{
\colhead{Num} & \colhead{SDSS Name} & \colhead{$z$} & \colhead{$\log(M_{\rm BH})$} & \colhead{$L_{\rm bol}/L_{\rm Edd}$} & \colhead{ObsID/SEQID} & \colhead{Count rate} &\colhead{Expo.} \\
\colhead{ } & \colhead{ } &\colhead{ } &\colhead{($M_{\odot}$)} &\colhead{ } &\colhead{ } &\colhead{(counts\,s$^{-1}$)} &\colhead{(ks)} & \colhead{${\sigma}_{\rm rms,10ks}^{2}$}  & \colhead{$Err_{\rm 10ks}$}  \\
\colhead{(1)} & \colhead{(2)} &\colhead{(3)} &\colhead{(4)} &\colhead{(5)} &\colhead{(6)} &\colhead{(7)} &\colhead{(8)} &\colhead{(9)} &\colhead{(10)}   \\
}

\startdata
X1  &  J102348.44+040553.7 & 0.099  & 5.44 & 0.32 & 0108670101  & 0.08 & 51.00  & 0.044 & 0.019/0.019    \\
    &                      &        &      &      & 0605540201  & 0.13 & 106.50 & 0.057 & 0.012/0.012    \\
X2  &  J114008.72+030711.4 & 0.081  & 5.70 & 0.90 & 0305920201  & 0.75 & 39.00  & 0.015 & 0.004/0.004    \\
X3  &  J122349.64+072657.9 & 0.075  & 5.63 & 0.55 & 0205090101  & 0.19 & 24.00  & 0.027 & 0.025/0.025    \\
X4  &  J143450.63+033842.6 & 0.028  & 5.73 & 0.06 & 0305920401  & 0.12 & 22.00  & 0.101 & 0.048/0.048    \\
    &                      &        &      &      & 0674810501  & 0.13 & 11.75  & 0.065 & 0.032/0.032    \\
X5  &  J010712.03+140845.0 & 0.077  & 6.09 & 0.34 & 0305920101  & 0.31 & 19.00  & 0.055 & 0.027/0.027    \\
X6  &  J135724.53+652505.9 & 0.106  & 6.20 & 0.47 & 0305920301  & 0.51 & 18.75  & 0.040 & 0.017/0.017    \\
    &                      &        &      &      & 0305920601  & 0.49 & 13.75  & 0.014 & 0.007/0.007    \\
X7  &  J082433.33+380013.2 & 0.103  & 6.11 & 0.41 & 0403760201  & 0.09 & 15.00  & 0.097 & 0.059/0.058    \\
X8  &  J091449.06+085321.1 & 0.140  & 6.28 & 0.37 & 0744220701  & 0.76 & 31.75  & 0.040 & 0.011/0.011    \\
X9  &  J134738.24+474301.9 & 0.064  & 5.63 & 0.67 & 0744220801  & 0.63 & 20.75  & 0.028 & 0.009/0.009    \\
X10 &  J115341.78+461242.3 & 0.025  & 6.09 & 0.29 & 0744220301  & 2.16 & 13.75  & 0.019 & 0.007/0.007    \\
R1  &  J122349.64+072657.9 & 0.075  & 5.63 & 0.55 & RP600009N00 & 0.10 & 16.00  & 0.186 & 0.077/0.076    \\
R2  &  J111644.65+402635.6 & 0.202  & 5.83 & 0.45 & RP700855N00 & 0.08 & 18.50  & 0.032 & 0.030/0.029    \\
\enddata
\tablecomments{Column 1: X or R denotes the object observed with {\it XMM-newton} or {\it ROSAT}, respectively, and X9, X10, X11 are new observations from our programme (proposal: 074422, PI: W. Yuan); Column 2: object name; Column 3: redshift; Column 4: black hole mass in units of the solar mass $M_{\odot}$, from \citet{D12}; Column 5: Eddington ratios; Column 6: observation ID for {\it XMM-Newton} or sequence ID for {\it ROSAT} observation; Column 7: mean count rate of each observation (counts\,s$^{-1}$); Column 8: cleaned exposure time (ks); Column 9 \& 10: excess variances and errors calculated on a timescale of 10\,ks.}
\end{deluxetable}



\section{MEASUREMENT OF EXCESS VARIANCE AND THE PSD MODELS}

\subsection{Excess Variance and the Uncertainty}

Following \citet{N97} (see also \citealp{T99,V03,P2004}), the normalized excess variance is calculated using the definition,
\begin{equation}
{\sigma}_{\rm rms}^{2}=\frac{1}{N{\mu }^{2}}\sum_{i=1}^{N}\left[\left( {X}_{i}-\mu \right)^{2} - \sigma_i^2\right],
\end{equation}
where $N$ is the number of good time bins of an X-ray light curve, $\mu$ the unweighted arithmetic mean of the count rates, $X_i$ and $\sigma_i$ the count rates and their uncertainties, respectively, in each bin.

As shown by \citet{V89, V97}, the excess variance is the integral of the PSD of a light curve over a frequency interval given by Eq.\,(2),
\begin{equation}
\sigma_{\rm rms}^2=\int_{\nu_{\rm min} }^{\nu_{\rm max}}P\left(\nu \right)d\nu,
\end{equation}
where $\nu_{\rm min}=\frac{1}{T}$, $\nu_{\rm max}=\frac{1}{2\Delta T}$, $T$ and $\Delta T$ are the time length and the binsize of the light curve, respectively. For a given light curve, it is clear that the exact value of the excess variance is dependent on the length of the light curve (e.g. observational duration) as well as on the binsize $\Delta T$. In order to compare the excess variances of different objects or observations in our sample, the duration (timescale) and the binsize of the light curves should be set to be the same for all the obejects. For this purpose, the light curves are divided into one or more segments of 10\,ks in length, and for each of the segment the excess variance is calculated. For observations having more than one segment, the mean of the segments is taken. The results are listed in Table \ref{tbl1}.

The uncertainty of the excess variance comes from two sources, one of the measurement uncertainty and the other of the stochastic nature of the variability process, as shown by \citet{V03}.
The measurement uncertainties of the excess variance are estimated following \citet{V03}:
\begin{equation}
\left(\Delta \sigma_{\rm rms}^2 \right)_{\rm mea}=\sqrt{\left(\sqrt{\frac{2}{N}}\frac{\left<\sigma_i^2\right>}{\mu^2}\right)^2+\left(\sqrt{\frac{\left<\sigma_i^2\right>}{N}}\frac{2F_{\rm var}}{\mu}\right)^2},
\end{equation}
where $\left<\sigma_i^2\right>$ is the mean of the square of count rate uncertainties, $F_{var}$ the fractional variability ($F_{\rm var}=\sqrt{\sigma_{\rm rms}^2}$), and the other quantities are defined in Eq.\,(1).

As discussed in \citet{V03}, the light curves of AGN are simply stochastic, meaning that each observed light curve is only one realization of the underlying random variability process, and each realization can exhibit a slightly different mean count rate and variance. The random fluctuations between different realizations lead to non-ignorable scatter in the excess variance. This phenomenon can be reflected in our result: for a source with more than one observation (such as J1023+0405, No. X1 in Table \ref{tbl1}), the excess variance varies considerably. The scatter can be significantly reduced if the light curves are sufficiently long, or, equivalently, having a sufficient number of data points.
We then compute the uncertainty owing to the stochastic nature of the variability process using the method introduced in \citet{V03}. We first build a PSD model \footnote{The PSD models used in this paper will be introduced in the next subsection, and all the models yield essentially the same results.} to simulate a light curve using the method of \citet{T95} with a binsize of 250\,s and sufficiently long duration. The light curve is divided into 1000 separate segments of 10\,ks. The distribution of the calculated excess variances of all the segments is obtained, from which the range of stochastic uncertainty is found. We take the 68\% confidence range for the expected excess variance on each of the timescales to get the stochastic scatter. Thus, from our simulations the scatter due to this random process is $\Delta\log(\sigma_{\rm rms}^2)=-0.29$ and $+0.24$ for 10\,ks light curve of a 250s-binsize.

Finally, the total uncertainties of the excess variance are obtained by combining in quadrature the stochastic and measurement uncertainties, which are at the $68\%$ confidence level. The obtained uncertainties of $\sigma_{\rm rms}^2$ are given in Table \ref{tbl1}.
For an object having more than one observation, the mean and its uncertainty are calculated and used as the final excess variance in the following analysis.

\subsection{PSD Models of AGN}

\begin{figure}
\epsscale{1.00}
\plotone{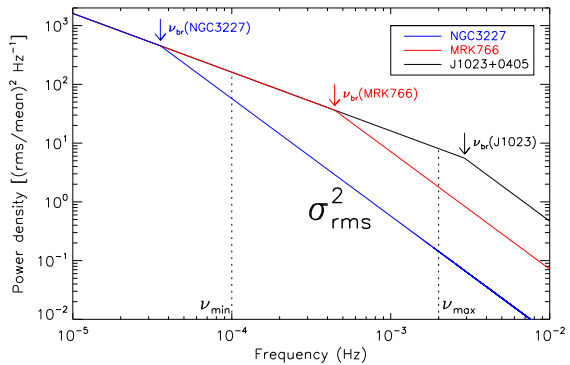}
\caption{Sketch of the power specturm density of AGN and its relationship with the excess variance, which is the integral of the PSD over the frequency span [$\nu_{\rm min}$, $\nu_{\rm max}$], where $\nu_{\rm max}$ is set by the binsize (250\,s) and $\nu_{\rm min}$ is set by the length (10\,ks) of the light curve. The PSD of three AGN with different $M_{\rm BH}$ and thus differnt break frequency $\nu_{\rm br}$ are shown as examples, which correspond to the three cases in Eq.\,(5) of $\nu_{\rm br}\leq\nu_{\rm min}$, $\nu_{\rm min}<\nu_{\rm br}\leq\nu_{\rm max}$, $\nu_{\rm br}>\nu_{\rm max}$, respectively. \label{fig3}}
\end{figure}

As mentioned above, the excess variance is the integral of the PSD of a light curve over a frequency interval. If the shape of the PSD is known, it is possible to calculate the values of the excess variances, and thus to derive the model relation between the excess variance and black hole mass, which can be compared to observations.
For AGN, it has been shown that the PSD function has a standard shape of a broken power-law,
\begin{equation}
P(\nu)=
\begin{cases}
A(\nu/\nu_{\rm br})^{-1}, & \nu \leq \nu_{\rm br}\\
A(\nu/\nu_{\rm br})^{-2}, & \nu>\nu_{\rm br}
\end{cases}
\end{equation}
where $\nu_{\rm br}$ is the break frequency \citep{M06,G12}. For an AGN with given $M_{\rm BH}$ and $\dot{m}$, $\nu_{\rm br}$ can be determined, though the exact value differs somewhat in different models. From Eq.\,(2) and (4) the excess variance can be expressed as
\begin{equation}
\sigma_{\rm rms}^2=
\begin{cases}
C_1\nu_{\rm br}(\nu_{\rm min}^{-1}-\nu_{\rm max}^{-1}), &  \nu_{\rm br}\leq\nu_{\rm min}\\
C_1\left[\ln\left(\frac{\nu_{\rm br}}{\nu_{\rm min}}\right)-\left(\frac{\nu_{\rm br}}{\nu_{\rm max}} \right)+1 \right], & \nu_{\rm min}<\nu_{\rm br}\leq\nu_{\rm max}\\
C_1\ln\left(\frac{\nu_{\rm max}}{\nu_{\rm min}}\right), & \nu_{\rm br}>\nu_{\rm max}
\end{cases}
\end{equation}
where $ C_1=A\nu_{\rm br}$ is defined as the PSD amplitude (\citealp{P04}, see also model A in \citealp{G11} for details).
It has been suggested that $\nu_{\rm br}$ is inversely scaled with $M_{\rm BH}$, with a possible dependence on the scaled accretion rate in the Eddington units. The relationship between PSD and $\sigma_{\rm rms}^2$ is illustrated in the sketch in Figure \ref{fig3}, in which the PSD of three objects in our sample or P12 are shown. The break frequencies derived from $M_{\rm BH}$ and $\dot{m}$ using the \citet{G12} scaling relation are also indicated for the three objects, which represent the three cases in Eq.\,(5), respectively.

However, although the broken power-law model given in Eq.\,(4) is widely accepted, the exact values of the amplitude $C_{1}$ and the break frequency $\nu_{\rm br}$ differ somewhat in different studies. \citet{P04} first suggested a model with $\nu_{\rm br}=17/(M_{\rm BH}/M_{\odot})$ (Hz) and $C_1=A\nu_{\rm br}=0.017$. Later studies with larger samples found a secondary dependence of $\nu_{br}$ on the scaled accretion rate $\dot{m}$ \citep{M06,G12}, in addition to the dependence on $M_{\rm BH}$. In the literatures of recent studies, three following PSD models of AGN were used:
\begin{description}
\item[Model A:] Adopted by \citet{G11}, this model assumes the scaling relation $\nu_{\rm br}=0.003\dot{m}(M_{\rm BH}/10^6\ M_{\odot})^{-1}$ suggested by \citet{M06}, and a constant PSD amplitude $C_{1}=A\nu_{\rm br}$ ($=0.017\pm 0.006$, \citealp{P04}).
\item[Model B:] The scaling relation for $\nu_{\rm br}$ is the same as in model A. As argued by \citet{P12}, the dispersion of the $M_{\rm BH}-\sigma_{\rm rms}^2$ relation arising from different $\dot{m}$ derived from model A is too large to account for the observational data, and a new $\dot{m}$ dependent PSD amplitude was suggested: $C_{1}= \alpha\dot{m}^{-\beta}$, where $\alpha=0.003^{+0.002}_{-0.001}$ and $\beta=0.8\pm0.15$ were fitted in their study.
\item[Model C:] Similar to model A ($C_{1}=0.017$), but an improved scaling relation $\nu_{\rm br}=0.001\dot{m}^{0.24}(M_{\rm BH}/10^6\ M_{\odot})^{-1}$ suggested by \citet{G12}, is adopted, which predicts a weaker dependence on $\dot{m}$.
\end{description}

\section{RESULTS AND DISCUSSION}

\subsection{$M_{\rm BH}-\sigma_{\rm rms}^2$ Relation in the Low-mass Regime}

The measured excess variances of 11 low-mass AGN of our sample are calculated from the X-ray light curves obtained in Section 3 using Eq.\,(1). The results are listed in Table \ref{tbl1}. To enlarge the sample, we make use of both the {\it XMM-Newton} and {\it ROSAT} data, which have somewhat different energy bands. It has been shown that the excess variance is not sensitive to energy bands in the range concerned here \citep{P12}. To enlarge the working sample we also include 8 objects \footnote{They are GH18, GH49, GH78, GH112, GH116, GH138, GH142, and GH211.} presented in \citet{L15} which are not in our sample. The excess variance values given in \citet{L15} are used, which were calculated using a binsize of 200\,s, different from ours (250\,s). However, the effect of such a difference on the calculated excess variance is negligible ($<2\%$), as found by performing simulations using the method of \citet{T95}.

\begin{figure}
\plotone{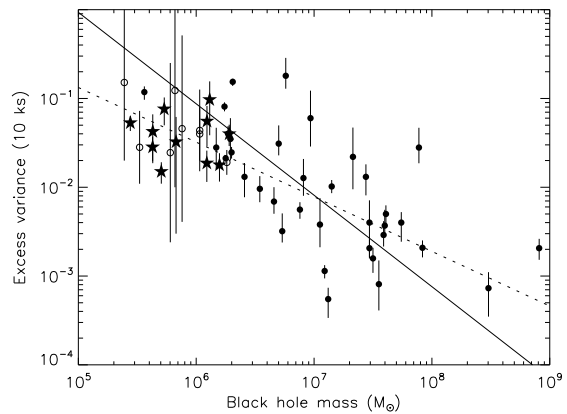}
\caption{Relationship of the black hole mass and the excess variance, which are measured on a timescale of 10\,ks, for our sample objects (stars), non-overlapping sources from \citet{L15} (open circles), and the sample of \citet{P12} (dots). The errors are at the 68\% confidence level. The solid line is the best-fit relation (slope $\approx$ -1) given by \citet{P12} based on that sample only, and the dotted line represents the best-fit relation (slope $\approx$ -0.58) for both the low-mass and the P12 sample (see text). \label{fig4}}
\end{figure}

Here we study the relation between $\sigma_{\rm rms}^2$ and $M_{\rm BH}$ by combining the low-mass sample and the P12 sample. The combined sample has $M_{\rm BH}$ spanning four orders of magnitude, $10^5- 10^9\ M_{\odot}$. The results are shown in Figure \ref{fig4}. Also plotted is the relation [$\log(\sigma_{\rm rms}^2)=-2.09-1.03\log(M_{\rm BH}/10^7 M_{\odot})$] derived from the P12 sample only by \citet{P12}. It shows that the $\sigma_{\rm rms}^2$ values of the low-mass sample are comparable to the largest values in P12 with higher $M_{\rm BH}$. However, for all the sources in the combined low-mass sample except one, the excess variances fall systematically below the extrapolation of the $M_{\rm BH}-\sigma_{\rm rms}^2$ relation derived from the high-mass P12 sample. Our result indicates that, in the low-mass regime, the $M_{\rm BH}-\sigma_{\rm rms}^2$ relation may deviate from the previously known linear relation, and is likely to flatten at around $\sim 10^6 M_{\odot}$ toward the low-mass end. In fact, the low-mass sample objects themselves do not show any correlation between $\sigma_{\rm rms}^2$ and $M_{\rm BH}$ (the Spearman's correlation test gives a null probability of 0.42). To quantify the statistical significance of this deviation, we perform two statistical tests, assuming that the previous inverse linear relation is a good description of the $M_{\rm BH}-\sigma_{\rm rms}^2$ relation over the entire mass range. First, we use the two-sided binomial test to find out the probability of having 18 out of 19 low-mass AGN fall below the relation as observed, whereas an equal probability (50\%) of falling on either side of the relation is expected. This gives a probability of $7\times10^{-5}$. Moreover, we fit an inverse linear relation with the slope fixed at -1 in the log-log space to the data of both the low-mass and the P12 sample. The reduced $\chi^2/dof$ of the fit is 836/53. Then we assume that the inverse linear relation breaks to a constant at masses below a so-called break mass, which is fitted as a free parameter. The $\chi^2/dof$ value for this model is 705/52 (the break mass is fitted to be $1.7\times10^6 M_{\odot}$). This improves the fit dramatically by reducing the $\chi^2$ by $\Delta \chi^2=131$ for one additional free parameter. Despite the fact that although the fitting is statistically not acceptable in terms of the large reduced $\chi^2$ (may arise from some intrinsic scatters inherent to the relationship), as an approximation, we employ the F-test \citep{B92} to test the significance of adding the flattening term does NOT improve the fit. This yields a small $p$-value 0.003.
We thus conclude that the flattening of the inverse linear $M_{\rm BH}-\sigma_{\rm rms}^2$ relation toward low-mass AGN is statistically significant. The fitted break mass is $\sim1.7\times10^6 M_{\odot}$.

We also fit a simple linear relation with the slope as a free parameter in the log-log space using the data of both the low-mass and the P12 samples, using the bisector method as in \citet{P12} (see also Appendix A of \citealp{B09}). The best-fit relations are over-plotted in Figure \ref{fig4} as dotted lines. We find $\log(\sigma_{\rm rms}^2)=(-2.09\pm0.044)+(-0.58\pm0.053)\log(M_{\rm BH}/10^7 M_{\odot})$. The slope $-0.58\pm0.053$ deviates significantly from the expected value -1  based on previous studies for AGN samples with higher $M_{\rm BH}$ (e.g. \citealp{Z10,P12}).

\begin{figure*}
\epsscale{2.0}
\plottwo{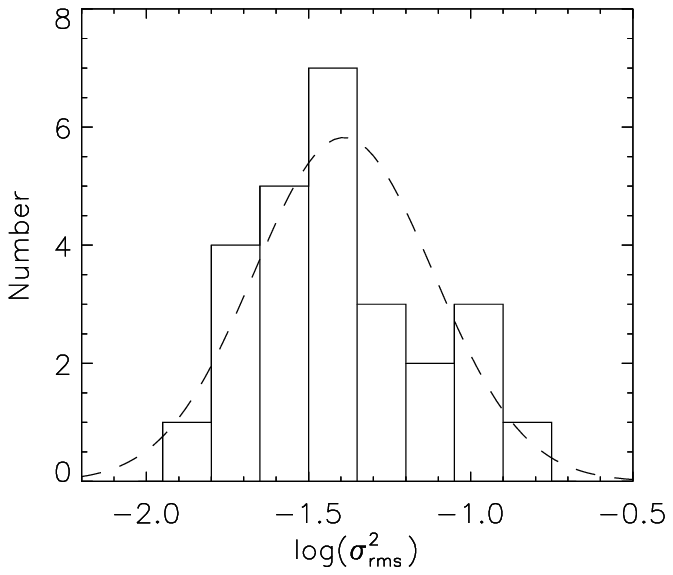}{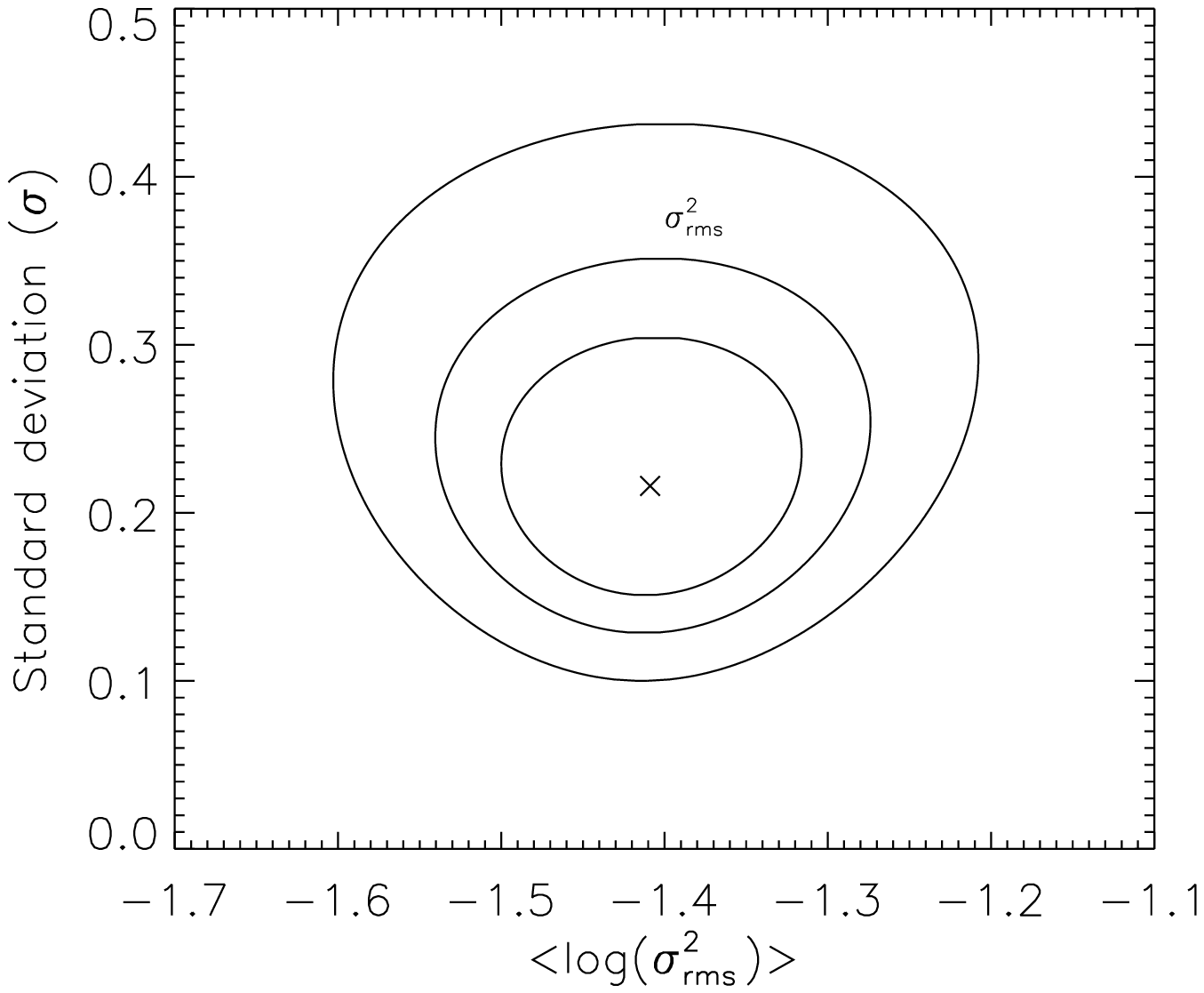}
\caption{Left panel: distributions of the excess variances for the objects with $M_{\rm BH}<2\times10^6M_{\odot}$. The dashed line represents a fitted Gaussian with a stardard deviation of 0.28. Right panel: confidence contours (at the 68\%, 90\% and 99\% confidence levels) of the mean and standard deviation of the intrinsic distribution of $\sigma_{\rm rms}^2$ (solid lines), which are derived using the maximum-likelihood method (see the text). The cross marks the best-estimated values. \label{fig5}}
\end{figure*}

For low-mass AGN, there appears to be a large scatter in the $\sigma_{\rm rms}^2$ values spanning almost one decade. The distribution of the excess variances (in logarithm) for our sample objects with $M_{\rm BH}<2\times10^6M_{\odot}$ is plotted in Figure \ref{fig5} (left panel), along with a fitted Gaussian distribution (dashed line). It would be interesting to examine the {\em intrinsic} dispersion of their $\sigma_{\rm rms}^2$ distribution, by taking into account the uncertainty of individual $\sigma_{\rm rms}^2$. The maximum-likelihood method as introduced by \citet{M88} is used to quantify the intrinsic distribution (assumed to be Gaussian) that is disentangled from the uncertainty of each of the measured excess variance (also assumed to follow a Gaussian distribution). We find a mean $\langle\log(\sigma_{\rm rms}^2)\rangle=-1.41^{+0.093}_{-0.091}$ with a standard deviation $\sigma=0.22^{+0.088}_{-0.065}$. The confidence contours of the two parameters are shown in Figure \ref{fig5} (right panel). This indicates that the apparently large scatter of $\sigma_{\rm rms}^2$ can mostly be attributed to the uncertainty of each of the measurements (including both the measurement and the stochastic uncertainties), and the intrinsic dispersion is small, but non-negligible (the standard deviation of $\log(\sigma_{\rm rms}^2)$ $\approx 0.15-0.3$ at the 68\% confidence level). It may also suggest that the dependence of $\sigma_{\rm rms}^2$ on any other parameters (e.g. accretion rate) is likely not strong.

\begin{deluxetable}{ccc}
\tabletypesize{\scriptsize}
\tablecaption{Fitting results of the $M_{\rm BH}-\sigma_{\rm rms}^2$ relation with the three PSD models.\label{tbl2}}
\tablewidth{0pt}
\tablehead{
\colhead{PSD model}  & \colhead{PSD amplitude} & \colhead{$\chi^{2}/dof$} \\
\colhead{(1)} & \colhead{(2)} &\colhead{(3)}  \\
}

\startdata
 A  & $C_1=0.032\pm0.001$                           & 2048/53  \\
 B  & $\alpha=0.0074\pm0.0004, \beta=0.80\pm0.02$   & 598/52   \\
 C  & $C_1=0.022\pm0.001$                           & 820/53   \\

\enddata
\end{deluxetable}

In conclusion, the previously found inverse scaling of $M_{\rm BH}-\sigma_{\rm rms}^2$ for AGN with supermassive black holes cannot be extrapolated to the low-mass regime, but starts to flatten at around $M_{\rm BH}\sim10^6M_{\odot}$. This implies that, although the excess variance can still be used to search for AGN with $M_{\rm BH}\lesssim10^6M_{\odot}$, it fails to provide accurate black hole mass estimation in this mass range. This result is consistent with that obtained in a recent paper by \citet{L15}. It is also suggested that, for AGN in the $10^5- 10^6\ M_{\odot}$ range, the intrinsic dispersion of the excess variances is likely to be small [standard deviation of $\log(\sigma_{\rm rms}^2)$ $\approx 0.15-0.3$].

\subsection{Comparison with Model Predictions}

Here we compare the observed $M_{\rm BH}-\sigma_{\rm rms}^2$ relation with the predictions based on the above three PSD models. In theory, the $M_{\rm BH}-\sigma_{\rm rms}^2$ relation can be deduced from the PSD of AGN using Eq.\,(5), once the shape and parameters of the PSD model are known. We take the three PSD models in Section 3.2, and fix the $\nu_{\rm br}(M_{\rm BH},\dot{m})$ relation as their original forms. Since the PSD amplitude $C_1$ was determined by fitting the $M_{\rm BH}-\sigma_{\rm rms}^2$ relation in previous work \citep{P04,P12}, it is set to be a free parameter here (two parameters $\alpha$ and $\beta$ for model B since $C_{1}= \alpha\dot{m}^{-\beta}$). The bivariate models of $\sigma_{\rm rms}^2(M_{\rm BH},\dot{m})$ from Eq.\,(5) are fitted to the data of both the low-mass and the P12 samples using the simple $\chi^2$-minimization fitting. The results are given in Table \ref{tbl2}, with the errors of the fitted normalization are at the $68\%$ confidence level. The best-fit PSD amplitudes are very close to their original values for all the models.

\begin{figure*}
\epsscale{2.0}
\plotone{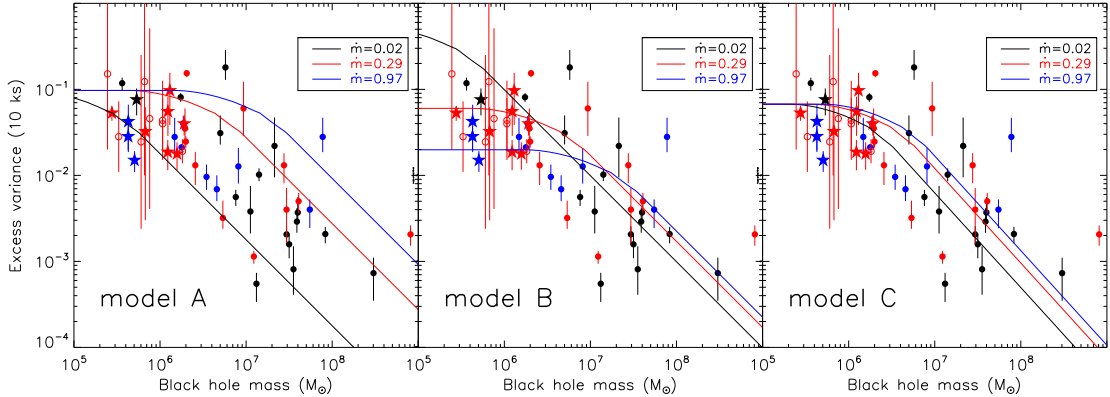}
\caption{The same as Figure \ref{fig4} except that the solid lines represent the theoretical relations derived from the best-fit PSD models assuming different normalization and the dependence of break frequency on black hole mass and accretion rate (model A, B, and C; see text for details). The three colors correspond to three accretion rates of $\dot{m}=$0.02 (black), 0.29 (red), 0.97 (blue), which are chosen to be the mean (in logarithm) of $\dot{m}$ in three $\dot{m}$ bins 0.01-0.1, 0.1-0.5, and 0.5-4.0 of the whole sample objects. The sources with $\dot{m}$ in the three $\dot{m}$ bins are plotted in corresponding colors. \label{fig6}}
\end{figure*}

The model $M_{\rm BH}-\sigma_{\rm rms}^2$ relations using the best-fit PSD normalizations are plotted in Figure \ref{fig6} for the three models, along with the data.
The black, red and blue solid lines represent the relations for three typical accretion rates ($\dot{m}=0.02, 0.29, 0.97$), respectively, which are the mean (in logarithm) of $\dot{m}$ in three $\dot{m}$ bins (0.01-0.1, 0.1-0.5, and 0.5-4.0). The objects with $\dot{m}$ in the three $\dot{m}$ bins are plotted in corresponding colors. It can be seen that all the three models can reproduce the observed trend of the $M_{\rm BH}-\sigma_{\rm rms}^2$ relation well quantitatively. For a given accretion rate, an inverse proportion is predicted in the high $M_{\rm BH}$ regime, whereas in the low-mass regime it flattens toward lower $M_{\rm BH}$. The exact value of $M_{\rm BH}$ at which the relation starts to flatten (the break mass) depends on $\dot{m}$: the higher $\dot{m}$, the larger the break mass. This trend generally holds for all the three PSD models, although the break mass varies from model to model, in addition to its dependence on $\dot{m}$.

Besides this general trend, there are some noticeable differences between the three models. For model A, the strong dependence on $\dot{m}$ leads to a large scatter in high-mass regime over a range of $\dot{m}$. Model B predicts a smaller scatter, but the scatter is relatively larger in the low-mass than in the high-mass regime. Compared to model A, model C also predicts a smaller scatter over the whole $M_{\rm BH}$ range, given its weak dependence of $\nu_{\rm br}$ on $\dot{m}$.

The inverse proportion of the $M_{\rm BH}-\sigma_{\rm rms}^2$ relation and its flattening toward low $M_{\rm BH}$ can easily be understood from Eq.\,(5) and Figure \ref{fig3}. For $\nu_{\rm br}\leq\nu_{\rm min}$ (e.g. NGC 3227 in Figure \ref{fig3}), as $\nu_{\rm br}$ increases (and $M_{\rm BH}$ decreases, since $\nu_{\rm br}\propto {M_{\rm BH}}^{-1}$), the integral of the PSD within [$\nu_{\rm min}$, $\nu_{\rm max}$] increases accordingly, resulting in a linearly increasing $\sigma_{\rm rms}^2$. Thus the previously known $M_{\rm BH}-\sigma_{\rm rms}^2$ relation (for $M_{\rm BH}>10^6M_{\odot}$) is a manifestation of the more fundamental dependence of $\nu_{\rm br}$ on $M_{\rm BH}$ (see also \citealp{P04,P12}). It also shows that the secondary dependence of $\nu_{\rm br}$ on $\dot{m}$ introduces  a scatter, the extent of which depends on the models. For $\nu_{\rm min}<\nu_{\rm br}\leq\nu_{\rm max}$ (e.g. MRK 766), the relation becomes non-linear. For $\nu_{\rm br}>\nu_{\rm max}$ (e.g. J1023+0405), $\sigma_{\rm rms}^2$ becomes independent of $\nu_{\rm br}$ (thus of $M_{\rm BH}$) and remains a constant.

We have shown that the observed $M_{\rm BH}-\sigma_{\rm rms}^2$ relation over almost the whole $M_{\rm BH}$ range for AGN of $M_{\rm BH}=10^5- 10^9\ M_{\odot}$ can be explained qualitatively by the PSD shape of AGN and the dependence of $\nu_{\rm br}$ on $M_{\rm BH}$, although the exact relation depends on the details of the models. It would be interesting to investigate which of the above models give a better description of the data, taking advantage of the extended $M_{\rm BH}$ range provided by our sample. Here we discuss this issue only briefly by comparing the deviations of the data from the models in terms of the fitted $\chi^2$, although they are too large for the fits to be acceptable {\em nominally} for all the three models (the large $\chi^2$ values are partly due to the somewhat large uncertainties in $M_{\rm BH}$ and $\dot{m}$, which are not taken into account in the fitting). We discuss the three models, respectively. (\romannumeral1) Model A has a much larger $\chi^2$ value than model B and C, and is likely not a good description of the data. The same suggestion was also argued by \citet{P12}. (\romannumeral2) For model B, The fitted parameter $\beta$ ($=0.80\pm0.02$) is consistent with that of \citet{P12}. For the objects with $M_{\rm BH} < 2\times10^6\ M_{\odot}$, no correlation is found between $\sigma_{\rm rms}^2$ and $\dot{m}$ (the Spearman correlation test gives a null probability $p=0.42$), which seems to be inconsistent to the model prediction (see Figure \ref{fig6}, middle panel). However, this may be partly due to the uncertainties in determining $\dot{m}$. (\romannumeral3) For model C, our result yields a PSD amplitude consistent with that of \citet{P04}, which is only weakly dependent on $\dot{m}$. This leads the excess variances converging to a constant value toward low $M_{\rm BH}$, which is roughly consistent with the small intrinsic scatter (not zero, however) found above. Overall, based on the fitted $\chi^2$ values (Table \ref{tbl2}), we tend to consider models B, and perhaps model C to be better descriptions of the PSD shape of AGN. However, a rigorous comparison of the PSD models is beyond the scope of this paper and will be carried out in future work with a larger sample and better quality of data.

\section{CONCLUSION}
The $M_{\rm BH}-\sigma_{\rm rms}^2$ relation for AGN in the low-mass regime ($M_{\rm BH}\lesssim10^6\ M_{\odot}$) is investigated using a sample of 11 low-mass AGN observed by {\it XMM-Newton} and {\it ROSAT}, including both new observations and archival data, as well as 8 sources from \citet{L15}.
We find that, the inverse linear $M_{\rm BH}-\sigma_{\rm rms}^2$ relation established in the high-mass regime in previous studies (e.g. \citealp{Z10,P12}) fails to extend to $M_{\rm BH}$ below $10^6\ M_{\odot}$. The relation becomes to flatten at $\sim10^6\ M_{\odot}$, below which the excess variances seem to remain constant. Our result is in good agreement with that obtained from a recent similar study by \citet{L15}. This is in fact consistent with the model prediction from our current understanding of the PSD of AGN and the dependence of the break frequency $\nu_{\rm br}$ on $M_{\rm BH}$. Our result suggests that while the X-ray excess variance may still be used to search for low-mass AGN candidates \footnote{As those already carried out by e.g. \citet{K12} and \citet{T12} by assuming that the reverse linear scaling relation can be extended to the low-mass regime}, it fails to provide reliable estimation of the black hole mass for AGN with $M_{\rm BH} \lesssim 10^6\ M_{\odot}$. In this $M_{\rm BH}$ regime, it is also found that the excess variances show small intrinsic dispersion, when their uncertainties (both measurement and stochastic) are taken into account.



\acknowledgments

This work is supported by the National Natural Science Foundation of China (Grant No.11473035, Grant No. 11033007 and Grant No. 11173029), and the Strategic Priority Research Program ``The Emergence of Cosmological Structures'' of the Chinese Academy of Sciences (Grant No. XDB09000000). This work is mainly based on observations obtained with XMM-Newton, an ESA science mission with instruments and contributions directly funded by ESA Member States and NASA. The {\it ROSAT} project is supported by the Bundesministerium f\"urBildung, Wissenschaft, Forschung und Technologie (BMBF) and the Max-Planck-Gesellschaft. This research has made use of the NASA/IPAC Extragalactic Database (NED) which is operated by the Jet Propulsion Laboratory, California Institute of Technology, under contract with the National Aeronautics and Space Administration.

\end{document}